# Fragility curves for power transmission towers in Odisha, India, based on observed damage during 2019 Cyclone Fani

Surender V Raj[a], Manish Kumar[a*] and Udit Bhatia[a]

[a] *Civil Engineering Discipline, Indian Institute of Technology Gandhinagar, Gandhinagar, India, 382355*

ABSTRACT

Lifeline infrastructure systems such as a power transmission network in coastal regions are vulnerable to strong winds generated during tropical cyclones. Understanding the fragility of individual towers is helpful in improving the resilience of such systems. Fragility curves have been developed in the past for some regions, but without considering relevant epistemic uncertainties. Further, risk and resilience studies are best performed using the fragility curves specific to a region. Such studies become particularly important, if the region is exposed to cyclones rather frequently. This paper presents the development of fragility curves for high-voltage power transmission towers in the state of Odisha, India, based on a macro-level damage data from 2019 cyclone Fani, which was obtained through concerned government offices. Two types of damages were identified, namely, collapse and partial damage. Accordingly, fragility curves for collapse and functionality disruption damage states were developed considering relevant aleatory and epistemic uncertainties. The latter class of uncertainties included that associated with wind speed estimation at a location and the finite sample uncertainty. The most significant contribution in the epistemic uncertainty was due to the wind speed estimation at a location. The median and logarithmic standard deviation for the 50$^{th}$ percentile fragility curve associated with collapse were close to that for the functionality disruption damage state. These curves also compared reasonably well with those reported for similar structures in other parts of the world.

## 1. Introduction

Lifeline infrastructure systems, for example, power transmission, transportation, and water distribution networks, are at times exposed to natural hazards such as tropical cyclones, earthquakes, and floods. The performance of infrastructure depends considerably on the performance of its individual components (e.g., [1]). The relationship between a hazard and the performance of a component is often described in terms of one or a series of fragility curve(s). These curves may present probabilities of failure of a component given a hazard level. Fragility curves have been developed for components in nuclear and other structures (e.g., [2], [3]), bridges (e.g., [4]–[6]), transmission towers (e.g., [7], [8]), and non-structural components (e.g., [9]). These curves are used to characterize the performance of a structure subjected to extreme loads, assess the risk to a component or a system under a spectrum of hazards (e.g., [10]), study the resilience of an infrastructure system (e.g., [11], [12]), and in financial decision making (e.g., [13]). This paper focuses on the development of fragility curves for lattice-type power transmission towers in India.

---

[*] Corresponding author.
 *E-mail address:* mkumar@iitgn.ac.in (M. Kumar).

Transmission towers are most vulnerable under strong winds, such as those generated during tropical cyclones. In the past, two broad strategies have been used to develop fragility curves for the lattice transmission towers. The first strategy considers damage data in the aftermath of an extreme event. This approach was used to develop the fragility curve for transmission towers in Texas, United States, based on the damages observed during hurricanes between 1999 and 2008 (e.g., [7]). The report did not consider epistemic uncertainties, for example, that associated with wind speed estimates at the location of towers. The second strategy is based on structural analysis, wherein a set of representative transmission towers is subjected to wind loads considering uncertainties in the material properties and loading. This approach has been used to develop fragility curves to study transmission towers in the United Kingdom and China (e.g., [1], [8]).

Fragility curves for transmission towers have been developed for specific regions. These curves do not consider relevant epistemic uncertainties, which is important for risk and resilience studies. Moreover, the importance of fragility curves being specific to a region in such studies cannot be overstated, particularly if the region (e.g., that considered in the present study) is frequently exposed to tropical cyclones. However, development of fragility curves for such regions is often challenging due to the lack of information on the properties of towers. This paper presents the development of fragility curves for high-voltage transmission towers in the state of Odisha, India, based on a macro-level damage data from 2019 cyclone Fani made available through government offices. The power transmission and distribution systems had received a financial loss of approximately INR 8,000 crores (USD 1.1 billion) during the cyclone [14], which is about 1.5% of the gross domestic product for the state. Two types of damage were identified in the towers, namely, collapse and partial damage. Monte-Carlo-based simulations and bootstrapping were used to account for the epistemic uncertainties associated with the estimation of wind speed at the location of towers and finite sample uncertainty. Bounding Engineering Demand Parameter (EDP) method (e.g., [15], [16]) was used to develop the fragility curves. The relative influence of the two types of epistemic uncertainties was studied. The fragility curves were compared with those reported from different parts of the world.

## 2. Damage data for high-voltage transmission towers during 2019 Cyclone Fani

Power transmission and distribution systems in the state of Odisha incurred heavy losses during the 2019 cyclone Fani. This paper focuses on the damage to high-voltage transmission towers (132 kV and 220 kV). Fig. 1 shows an image of a collapsed transmission tower in Puri, which is a coastal city in the state of Odisha. The data of damaged transmission towers were obtained through a Right to Information Act (RTI) application to Odisha Power Transmission Corporation Limited (OPTCL) (response dated October 29, 2019). The information included the location (geospatial location or the location number in the transmission corridor), voltage, height, and the level of damage. Damages in towers were observed in Khordha – Puri (132 kV), Nimapara – Samangara (132 kV), Chandaka – Nimapara (132 kV), Chowdwar – Bidanasi (132 kV), Pandiabili – Samngara (220 kV), Atri – Narendrapur (220 kV) and Chandaka – Mendhasal (220 kV) corridors, which span through the districts of Puri, Khordha and Cuttack. A total of 87 towers collapsed during cyclone Fani, while another 41 suffered partial damages (e.g., damage to the peaks, cross-arm and broken members) according to the RTI response. Heights of the damaged towers varied between 21 m and 50 m. In general, 220 kV towers were taller than 132 kV towers. Fig. 2 presents the histogram for heights of the damaged towers.

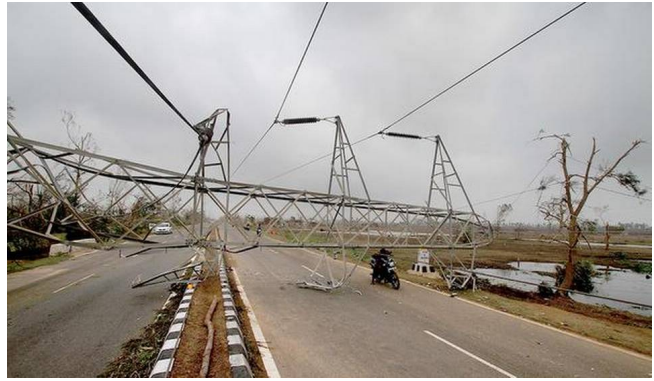

**Fig. 1.** A collapsed tower in Puri district of Odisha during 2019 cyclone Fani (Photo Credit: Biswaranjan Rout; taken from www.thehindu.com, accessed March 10, 2020.

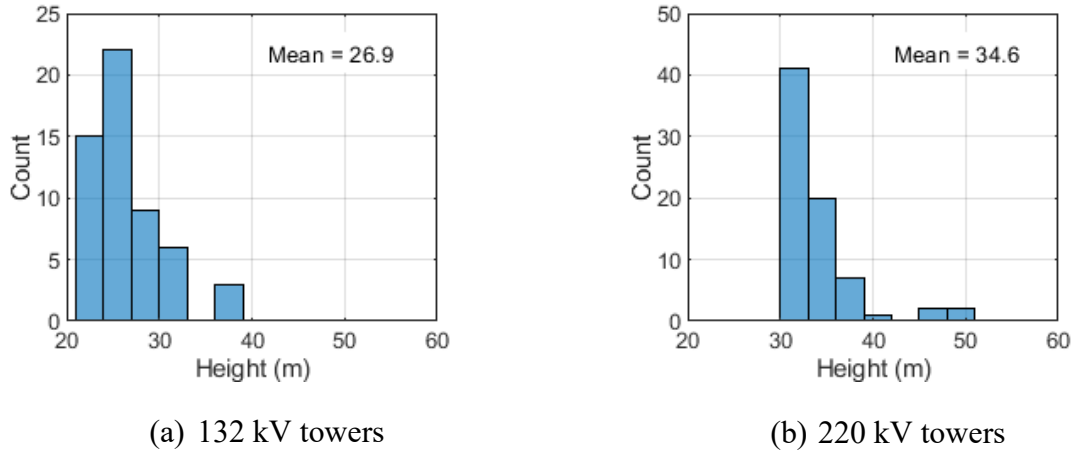

(a) 132 kV towers          (b) 220 kV towers

**Fig. 2.** Histograms for heights of towers damaged during 2019 cyclone Fani.

Fig. 3 shows the geospatial location of all 41,814 transmission towers within the geographical boundaries of the state of Odisha, which was obtained from Open Street Maps dataset (https://www.openstreetmap.org, accessed March 10, 2021). Location of the damaged transmission towers was identified either by mapping the geospatial location or the location number in the transmission corridor. The locations of the damaged transmission towers along with the trajectory of Cyclone Fani are shown in Fig. 4. The trajectory of the cyclone was obtained from the Indian Meteorological Department (IMD) website (http://www.rsmcnewdelhi.imd.gov.in, accessed March 10, 2021).

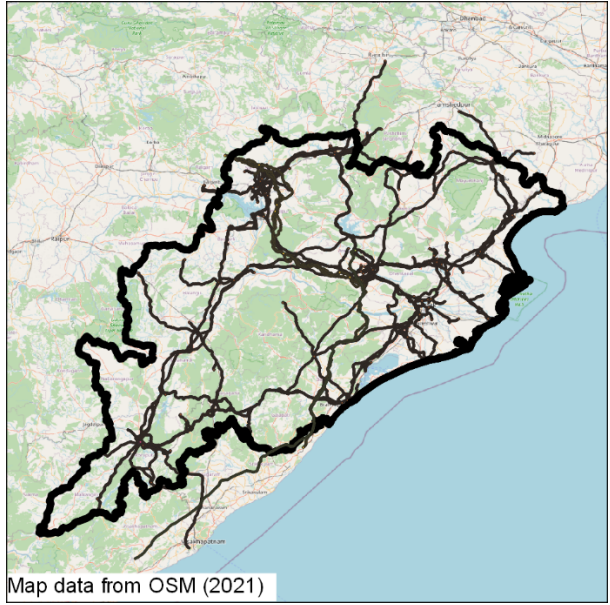

**Fig. 3.** Locations of all high-voltage transmission towers in the state of Odisha.

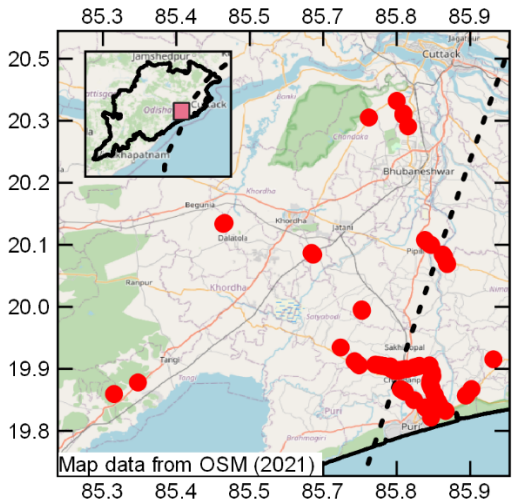

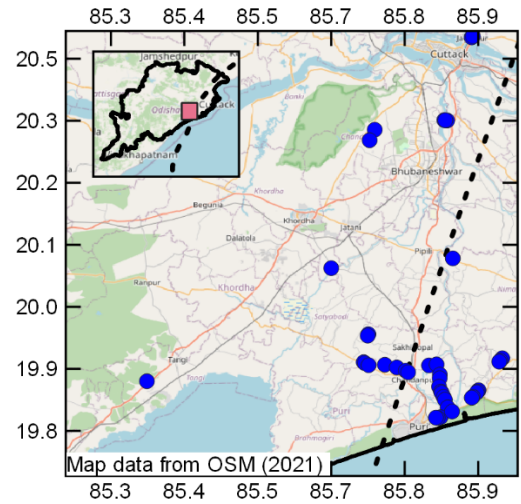

(a) Collapsed towers  (b) Partially damaged towers

**Fig. 4.** Locations of collapsed and partially damaged towers, and trajectory of cyclone Fani.

## 3. Method to develop fragility curves

Fragility functions present the probabilities of a structure (e.g., transmission tower) exceeding a particular level of damage for a given intensity measure. In this study, the 3-sec gust speed at a height of 10 m above the ground surface is selected as the intensity measure, which is consistent with relevant Indian standards (e.g., [17]). The gust speeds during cyclone Fani at the locations of 41,814 towers were estimated using available radial wind profile for tropical cyclones (e.g., [18], [19]). Two damage states were considered: (1) collapse, and (2) functionality disruption. As noted previously, 87 towers collapsed, while 128 towers (= 87+41) were in the functionality disruption damage state in the aftermath of Cyclone Fani. The choice of damage states (and corresponding fragility curves) in a study may depend on the task at hand. Functionality disruption fragility curves can be used to perform power transmission outage studies (e.g., [11], [20]), whereas fragility curves for both damage states may be useful in loss estimation (e.g., [13]).

Bounding Engineering Demand Parameter (EDP) method suggested by Porter et al. (e.g., [15], [16]) is used in this study to develop the fragility curves. All towers in the network are divided into $n$ bins with identical size based on their 3-sec gust speed estimates. Two parameters are calculated for each bin: (1) average 3-sec gust speed, and (2) ratio of the number of towers that reached the damage state of interest to the total number of transmission towers. Parameters of a lognormal cumulative distribution function are determined for this data using least square approach. Fragility curve is then given as follows (e.g., [15]):

$$F_{dm}(im) = \phi\left[\frac{\ln(im/X_{mdm})}{\beta_{dm}}\right] \qquad (1)$$

where, $\phi$ is standard normal cumulative distribution function, $F_{dm}$ is probability of exceedance for a damage state $dm$, $im$ is intensity measure, and $X_{mdm}$ and $\beta_{dm}$ are median and logarithmic standard deviation corresponding to the fragility curve, respectively. With the wind speeds at the locations of towers estimated using the Willoughby double exponential model [18], the median and the logarithmic standard deviation converged after $n = 30$ (see Fig. 5). Accordingly, $n$ for the present study was fixed at 30 for the present study.

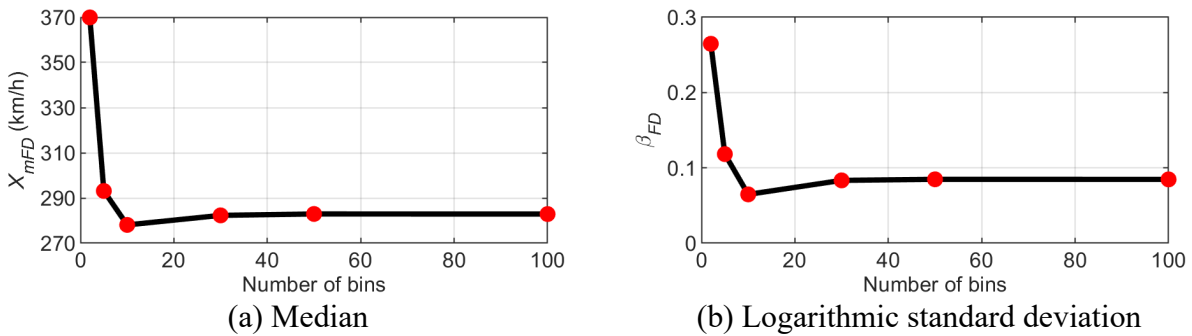

(a) Median  (b) Logarithmic standard deviation

**Fig. 5.** Fragility function parameters for functionality disruption damage state as a function of the number of bins.

## 4. Uncertainties considered in fragility curves

Fragility curves consider aleatory and epistemic uncertainties associated with the input parameters [21]. For example, the uncertainties in the geometrical and material properties of the towers would come under the former category, while that corresponding to the estimation of wind speed at a location would be considered as the latter. The aleatory uncertainty is accounted for through the logarithmic standard deviation $\beta_{dm}$[1]. Epistemic uncertainty reflects the lack of understanding of the phenomenon and is considered through multiple sets of models and associated input parameters with weights assigned to the set based on the judgment of the analyst. In the present study, epistemic uncertainties associated with the (1) estimate of intensity measure, and (2) size of sample are considered (e.g., [21], [22]). Details on the epistemic uncertainties considered are presented in the following sections. Corresponding results for both the damage states are presented in Section 5.

### 4.1. Estimation of 3-sec gust speed

As noted previously, the trajectory and wind speed data of cyclone Fani were obtained from the IMD's website. The intensity of the cyclone was available in terms of 3-min sustained wind speed at 10 m level from the ground surface. Fig. 6 shows the 3-minute sustained wind speed and the 3-sec gust for a wind speed history. The gust factor (GF) is defined as the ratio of the 3-sec to 3-min wind speeds. World Metrological Organization (WMO) recommends a GF of 1.58 with a standard deviation of 0.1 [23]; the associated distribution is assumed normal in the present study.

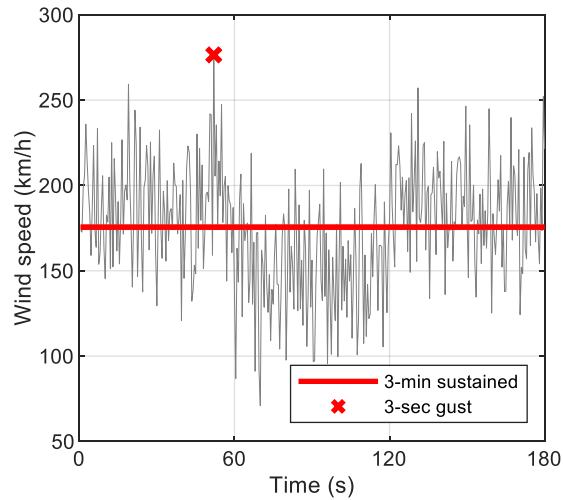

**Fig. 6.** Application of gust factor to convert 3-min to 3-sec wind speed.

Exact measurements of wind speed at all towers are often not available because either the device is not installed at the location of the tower or it is not functional. Therefore, radial wind profile models (RWPMs) are commonly used to estimate wind speed at the locations of interest. Three RWPMs are considered in the present study: Willoughby single exponential (WSE) [18],

---
[1] As will be discussed later in the paper, the fragility curves are defined using the parameters of a lognormal distribution.

Willoughby double exponential (WDE) [18], and Holland (HOL) [19]. A brief description of these models is presented in Appendix A[2]. Fig. 7 presents a comparison of radial wind profiles during 2019 cyclone Fani obtained using the three models at the landfall and 6 hours after the landfall.

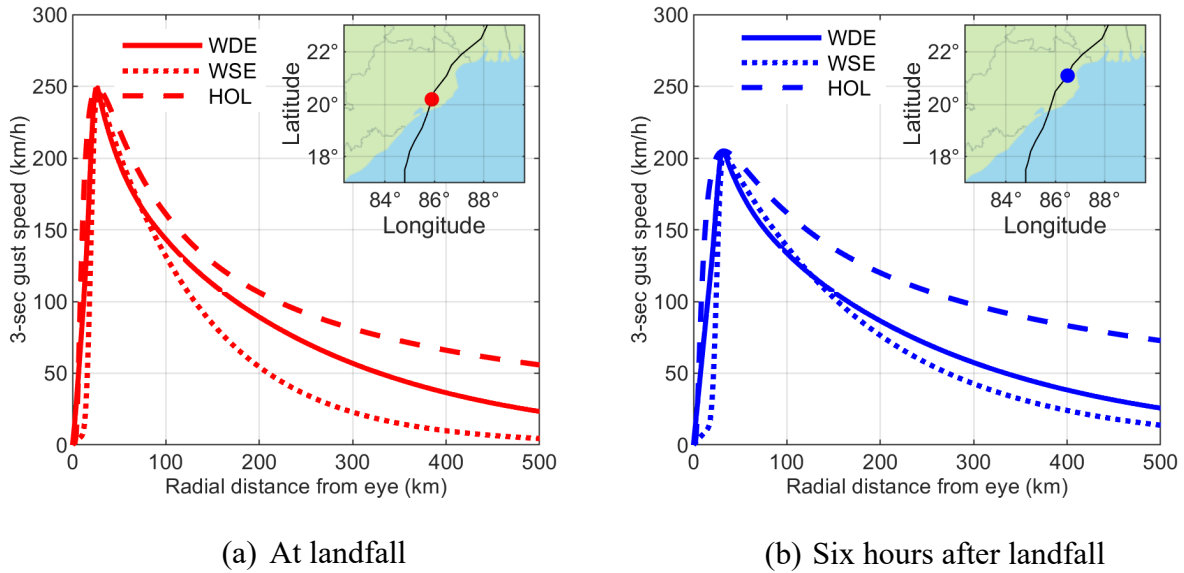

(a) At landfall  (b) Six hours after landfall

**Fig. 7.** Radial wind profiles estimated using WDE, WSE and HOL models during 2019 cyclone Fani.

Inputs provided to an RWPM include location of the eye of the cyclone with time and associated intensity in terms of maximum wind speed at gradient level[3]. The gradient level wind speed is obtained by dividing the 3-sec gust speed at 10 m height ($V_{10}$) by a conversion factor (*CF*). Three factors are considered in the present study: 0.75, 0.80 and 0.90 (see [24]). Fig. 8 shows the gradient levels corresponding to the three *CF*s.

One thousand sets of three random numbers were generated using Latin Hypercube (LH) sampling technique [22]. The three random numbers in a set correspond to three parameters: (1) RWPM, (2) *CF*, and (3) *GF*. For each set of three random numbers, RWPM, *CF* and *GF* were determined using the criteria presented in Eqs. (2), (3) and (4), respectively, which was used to calculate the wind speed at the locations of each of the 41,814 towers. At the end of this exercise, there were 1,000 estimates of wind speeds for the 41,814 towers during 2019 cyclone Fani.

---

[2] Steps involved in the estimation of wind speed at the desired site locations are described in the *stormwindmodel* package of R language (https://cran.r-project.org/web/packages/stormwindmodel, accessed June 2020).
[3] Gradient level is the height from ground surface at which the effect of surface friction can be neglected [25].

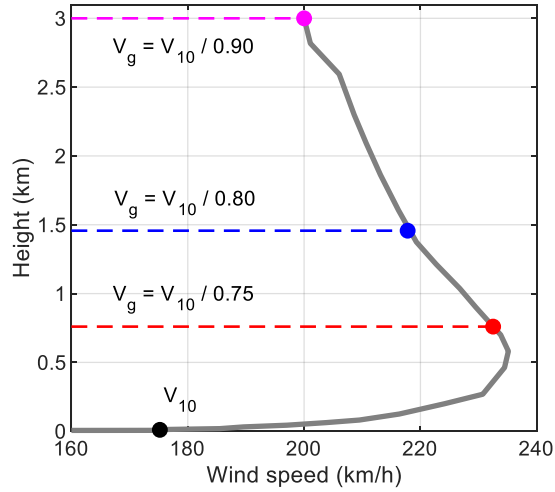

**Fig. 8.** Typical profile of wind speed at the eye walls during tropical cyclones.

$$\text{RWPM} = \begin{cases} WSE, & \text{if } 0 < r_1 \le \frac{1}{3} \\ WDE, & \text{if } \frac{1}{3} < r_1 \le \frac{2}{3} \\ HOL, & \text{if } \frac{2}{3} < r_1 \le 1 \end{cases} \qquad (2)$$

$$CF = \begin{cases} 0.75, & \text{if } 0 < r_2 \le \frac{1}{3} \\ 0.80, & \text{if } \frac{1}{3} < r_2 \le \frac{2}{3} \\ 0.90, & \text{if } \frac{2}{3} < r_2 \le 1 \end{cases} \qquad (2)$$

$$GF = \phi^{-1}(r_3 \mid 1.58, 0.1) \qquad (3)$$

where, $r_1$, $r_2$, and $r_3$ represents a set of three random numbers, and other parameters were defined previously. Fig. 9(a) presents the histogram of 1,000 estimates of wind speed at a tower location. Corresponding empirical cumulative distribution function is shown in Fig. 9(b). Median and the 68% confidence intervals of the 3-sec gust speed estimates at all the transmission tower locations are indicated in Fig. 10.

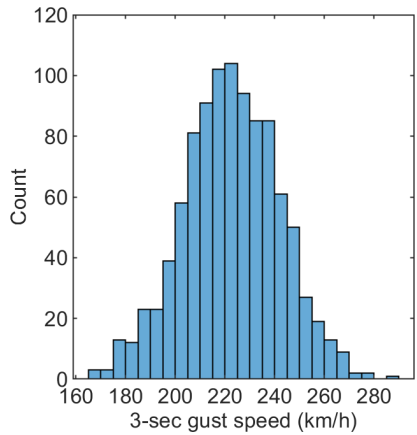 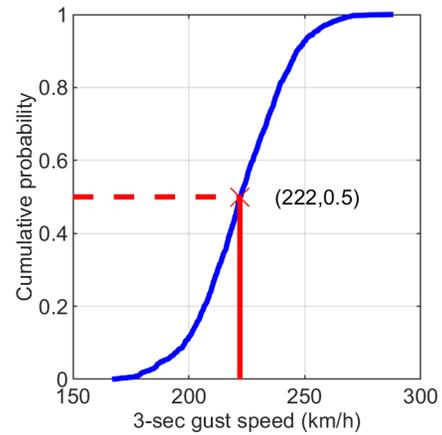

(a) Histogram of 3-sec gust speed estimates at a transmission tower location

(b) Empirical cumulative distribution function for 3-sec gust speed estimates

**Fig. 9.** Estimating 3-sec gust speed at a transmission tower location.

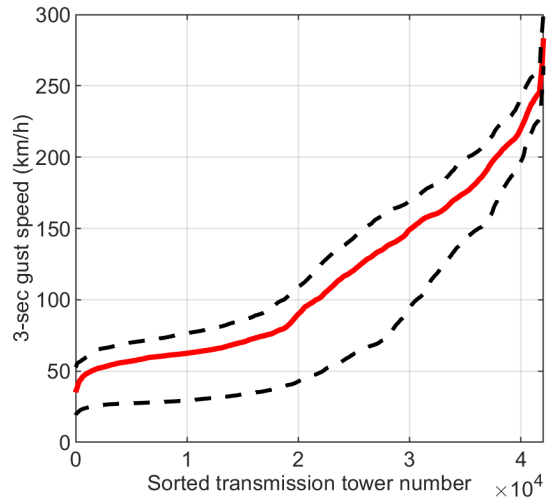

**Fig. 10.** Median and the 68% confidence bound estimates of the 3-sec gust speed at all the transmission tower locations (plotted after sorting) during Cyclone Fani.

Next, a series of Monte-Carlo simulations was performed (e.g., [22]). A random number between 0 and 1 was generated. For a given tower, this number multiplied by 100 was considered as a percentile, and the wind speed for the tower was obtained using the corresponding cumulative distribution function for the wind speed (e.g., Fig. 9(b)). The wind speed was estimated for all 41,814 towers in a similar manner. These wind speed estimates were used to develop a fragility curve for the transmission towers in the state of Odisha using the Bounding EDP method (see Section 3). The above exercise was repeated 1,000 times leading to fragility curves that consider the uncertainty associated with the wind speed estimation.

*4.2. Finite sample uncertainty*

Fig. 11 shows the 132 kV, 220 kV and 400 kV transmission lines near the coast of Odisha. Also shown in the figure is the track of cyclone Fani. It is clear that the 132 kV and 220 kV lines were most exposed to the cyclone, and therefore the damage was expected primarily in these towers. This is also consistent with the damage data provided by the OPTCL. However, it is possible that the response provided by the OPTCL missed out some damaged towers, including some 400 kV towers[4]. In view of the above, it becomes important to consider the finite sample uncertainty, which arises on account of consideration of only a sample of the entire damaged population. A common method to include the uncertainty is bootstrapping (e.g., [12], [13]). A new dataset of size identical to the original dataset is drawn from the original dataset with replacement. In the present study, size of the dataset is 41,814. It is possible in the process that some entries are repeated, while some others are not included. Radial wind profile model, *CF* and *GF* considered to assess the finite sample uncertainty were Willoughby double exponential model, 0.9 and 1.58, respectively. The new dataset was considered to estimate the fragility function parameters using the Bounding EDP method (see Section 3). This process was repeated 1,000 times (e.g., [21]).

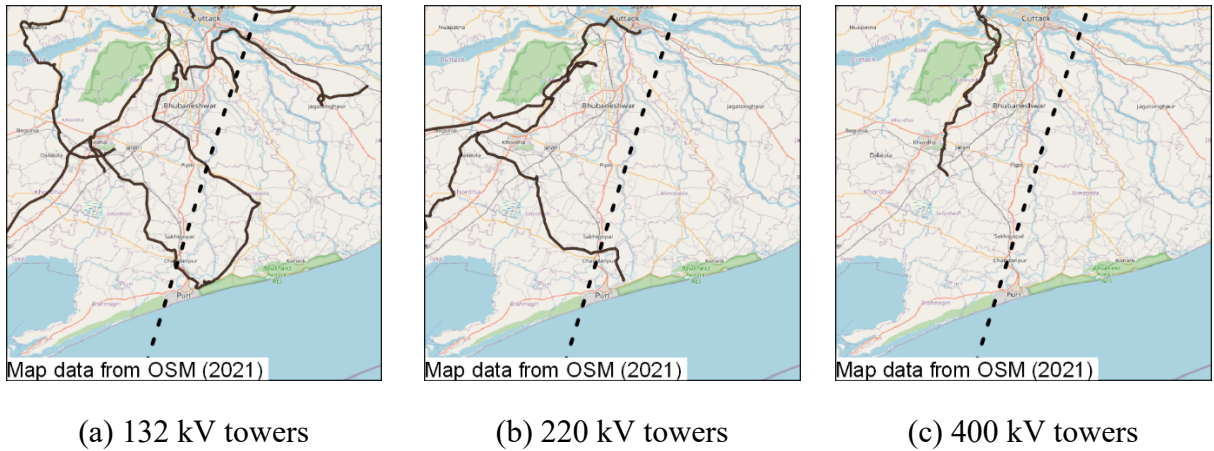

(a) 132 kV towers  (b) 220 kV towers  (c) 400 kV towers

**Fig. 11.** Locations of 132 kV, 220 kV and 400 kV transmission towers near the cyclone track of Fani.

*4.3. Combining epistemic uncertainties*

Fig. 12 presents the framework to combine the uncertainties in the intensity measure estimate and finite sample uncertainty (e.g., [21], [22]). One thousand random numbers between 0 and 1 were generated first. For a given random number, wind speeds at all 41,814 towers' locations were estimated using the respective empirical cumulative density function (e.g., Fig. 9(b)). This dataset of size 41,814 was resampled with replacement to yield another dataset of same size. The Bounding EDP method (see Section 3) was used to generate the fragility curve. The process was repeated for each of the 1,000 random numbers. Resulting fragility curves are considered to include the uncertainties in the intensity measure estimation and finite sample uncertainty.

---

[4] A number of RTI applications were filed to obtain the data considered in the present study. Response (dated September 18, 2019) to one of these applications indicated that a 400 kV tower was also damaged. However, additional details were not available for this tower. The response received on a later date (and considered in the present study) indicated that only 132 kV and 220 kV towers were damaged during 2019 cyclone Fani.

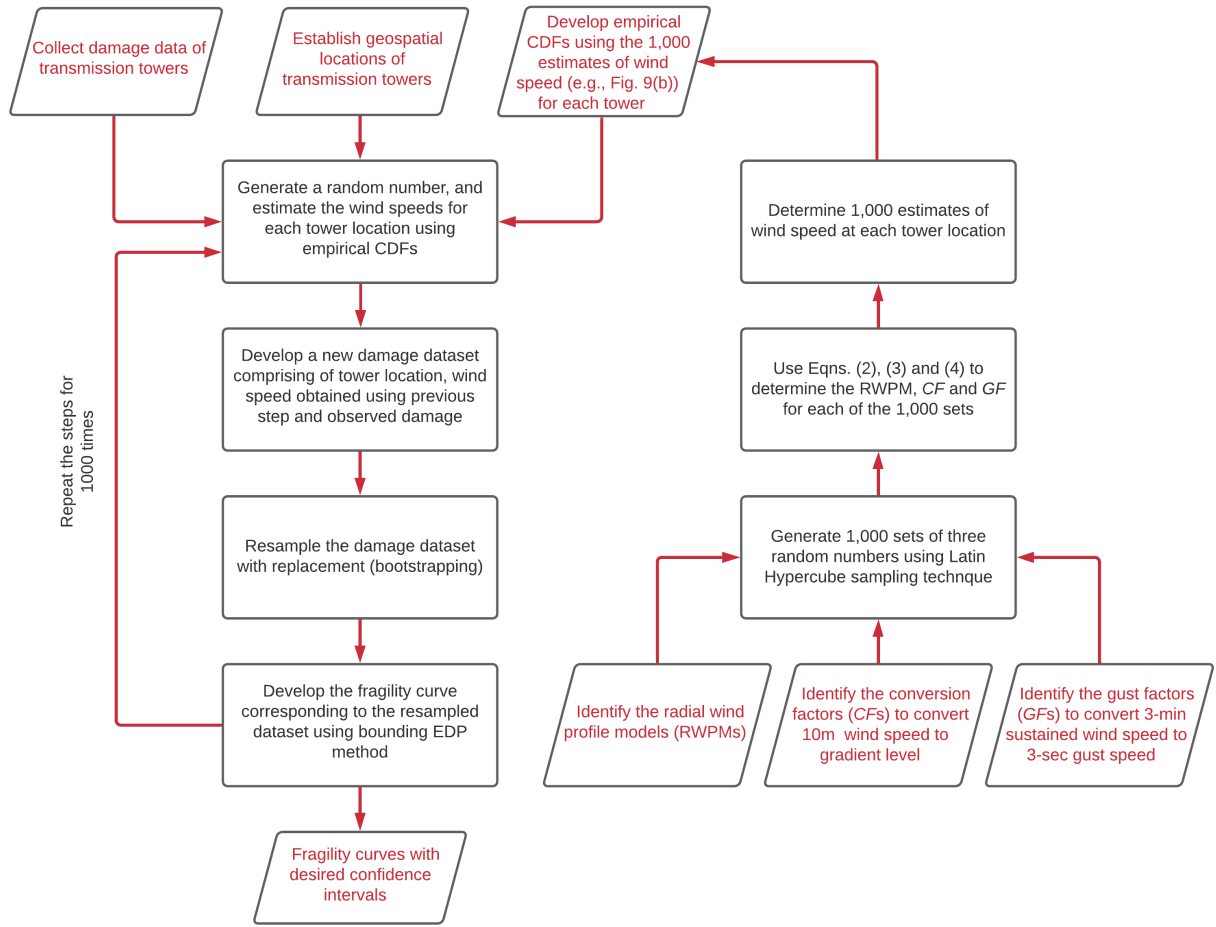

**Fig. 12.** Framework to combine uncertainties associated with intensity measure and sample size.

## 5. Results and discussions

Fig. 13(a) presents the fragility functions corresponding to functionality disruption damage state considering the aleatory uncertainty, and epistemic uncertainty associated with estimation in the intensity measure. As noted previously, the aleatory uncertainties are considered through a lognormal distribution. A discussion on the suitability of the assumed distribution is presented in Appendix B. Fragility curves corresponding to $16^{th}$, $50^{th}$ and $84^{th}$ percentiles are also shown in the figure. Medians of these fragility curves are 300.5 km/h, 279.9 km/h and 261.7 km/h, respectively. Fig. 13(b) presents the fragility curves considering aleatory, and finite sample uncertainties. The medians of the fragility curves corresponding to the three percentiles are 285.7 km/h, 282.4 km/h and 279.3 km/h, respectively, which indicates that the epistemic uncertainties are dominated by the intensity measure. This is also reflected in the fragility curves considering aleatory and both the epistemic uncertainties, as shown in Fig. 13(c). The medians corresponding to $16^{th}$, $50^{th}$ and $84^{th}$ percentile fragility curves are 299.9 km/h, 280.4 km/h and 260.8 km/h, which are rather close to that in Fig. 13(a). The logarithmic standard deviations associated with the three percentiles are 0.081, 0.089 and 0.095, respectively, in Fig. 13(a). These parameters also are very similar to those for Fig. 13(c).

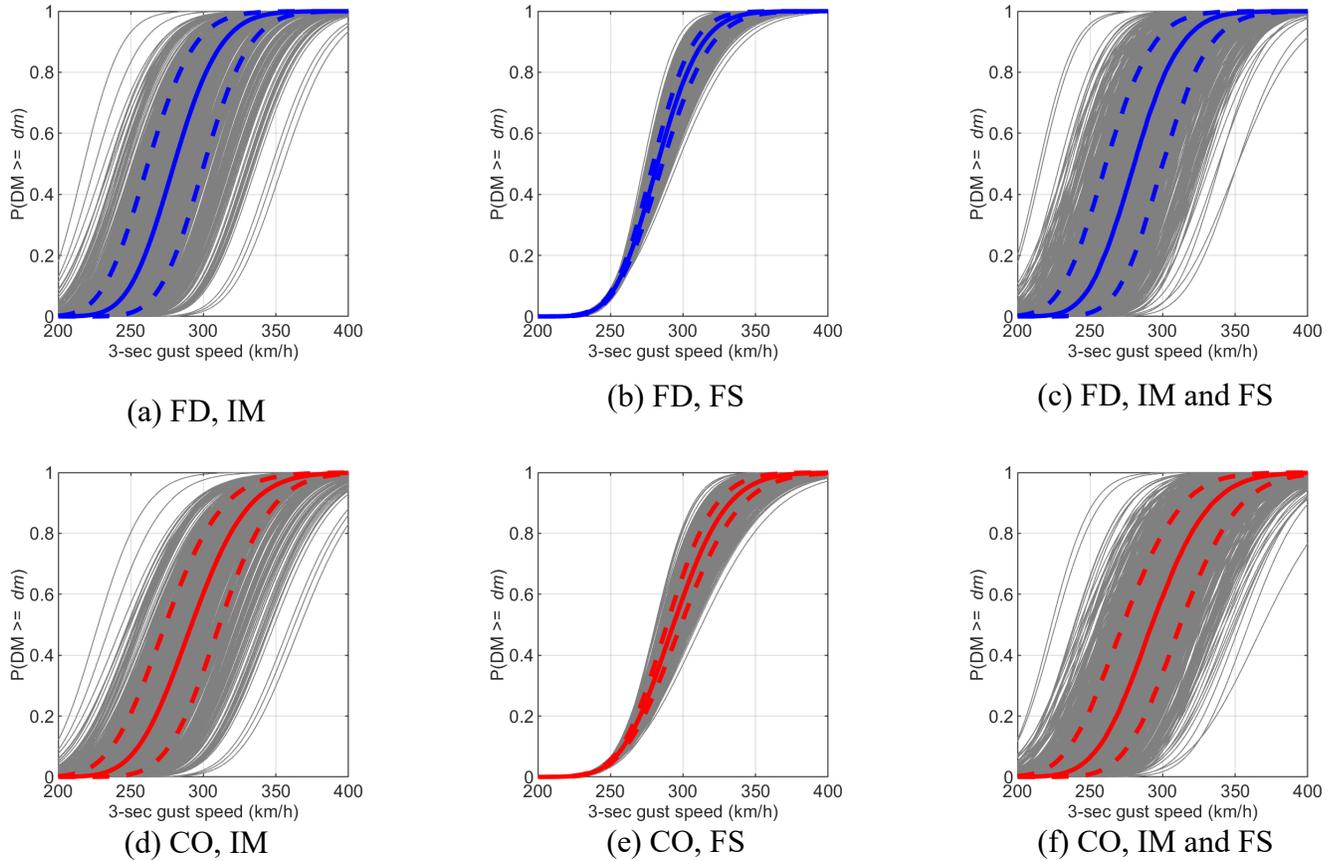

**Fig. 13.** Fragility functions for the two damage states (1) Collapse (CO) and (2) Functionality disruption (FD) considering intensity measure (IM), finite sampling (FS), and combined uncertainties.

Fig. 13(d) through Fig. 13(f) present the results corresponding to the state of collapse. General observations mentioned for the functionality disruption damage state are also applicable to the collapse state. Expectedly, the medians of fragility curve corresponding to collapse are marginally greater compared to those for functionality disruption. As a comparison, the median for 50$^{th}$ percentile fragility curve are 279.9 km/h and 291.9 km/h in Fig. 13(a) and Fig. 13(d), respectively. Table 1 summarizes the medians and logarithmic standard deviations corresponding to different fragility curves shown in Fig. 13. The results presented in Fig. 13(c) and Fig. 13(f) were generated for a new set of 1,000 random numbers between 0 and 1 (see Section 4.3). The new results were very similar to those presented in Fig. 13. As an example, the median of the 50$^{th}$ percentile fragility curve considering both the uncertainties was 293.3 km/h for the collapse damage state, which is very close to the corresponding value in Table 1 (292.5 km/h).

**Table 1.** The parameters of the lognormal fragility functions for collapse and functionality disruption damage states considering intensity measure and/or finite sample uncertainties.

| Uncertainty | Percentile (%) | Lognormal fragility function parameters | | | |
|---|---|---|---|---|---|
| | | Collapse | | Functionality disruption | |
| | | $X_{mSC}$ | $\beta_{SC}$ | $X_{mFD}$ | $\beta_{FD}$ |
| Intensity measure | 2.5 | 334.5 | 0.091 | 323.3 | 0.072 |
| | 16 | 311.1 | 0.092 | 300.5 | 0.081 |
| | **50** | **291.9** | **0.103** | **279.9** | **0.089** |
| | 84 | 273.8 | 0.111 | 261.7 | 0.095 |
| | 97.5 | 256.3 | 0.122 | 244.1 | 0.106 |
| Finite sample | 2.5 | 304.4 | 0.115 | 289.7 | 0.097 |
| | 16 | 298.5 | 0.105 | 285.7 | 0.091 |
| | **50** | **293.4** | **0.096** | **282.4** | **0.083** |
| | 84 | 288.9 | 0.087 | 279.3 | 0.076 |
| | 97.5 | 285.6 | 0.083 | 276.5 | 0.070 |
| Combined | 2.5 | 334.6 | 0.095 | 323.3 | 0.083 |
| | 16 | 312.0 | 0.100 | 299.9 | 0.085 |
| | **50** | **292.5** | **0.104** | **280.4** | **0.088** |
| | 84 | 272.8 | 0.112 | 260.8 | 0.096 |
| | 97.5 | 254.6 | 0.113 | 241.6 | 0.108 |

The median fragility curves shown in Fig. 13(c) and Fig. 13(f) are reproduced in Fig. 14. Also shown in Fig. 14 are fragility curves for electrical transmission towers reported from different parts of the world. The fragility curve in Quanta report [7] was developed based on the damage data of transmission structures during hurricanes that hit the state of Texas in the United States between 1999 and 2008. Fragility curves proposed by Pantelli et al. [1] and Fu et al. [8] were developed based on structural analysis on representative towers from United Kingdom and China, respectively. The former study considered aleatory uncertainty associated with wind loads, while the latter considered uncertainties associated with both wind loading and material properties. The 50[th] percentile fragility curves for collapse and functionality disruption developed in the present study compare reasonably with the corresponding fragility curves in other parts of the world. Towers considered by Fu et al. [8] appear to represent regions of rather low wind speed than that considered in the present study. Table 2 presents the parameters of the fragility functions shown in Fig. 14.

The fragility curves for the transmission towers in Odisha, India indicate the damage to these towers would not take place until the 3-sec gust speed of 220 km/h. Further, all towers are expected to lose functionality (collapse) at the wind speeds of 350 km/h (375 km/h) or greater. Probability of collapse is smaller than the probability of functionality disruption for a given wind speed, which is expected. However, there are exceptions. At low wind speeds (e.g., 220 km/h or lower), the probability of collapse is greater than that for functionality disruption. This is attributed to a greater logarithmic standard deviation associated with the collapse fragility curve (also see Table 1). Such observations have been made in the past as well, and recommendation to resolve the same has also been made (e.g., [15]). As an example, the fragility curves can be redefined such that the probability associated with the functionality disruption damage state is always greater than that for collapse.

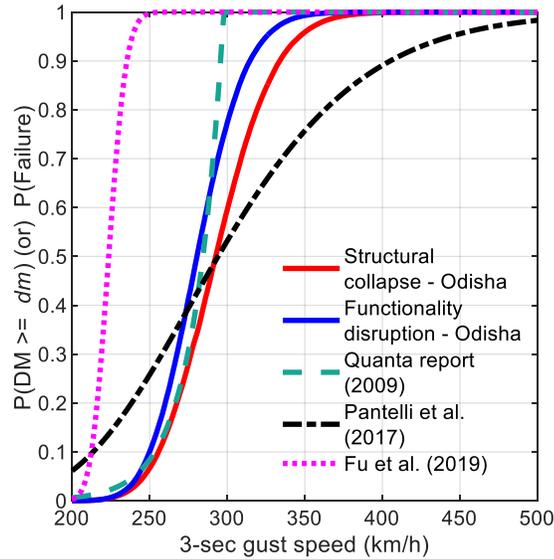

**Fig. 14.** 50th percentile fragility curves of structural collapse and functionality disruption damage state compared with fragility functions from previous literature for transmission towers in other parts of the world.

**Table 2.** The parameters of the developed fragility functions compared with the existing fragility functions for the transmission towers.

| Sl. No | Source | Parameters | |
|---|---|---|---|
| | | $x_m$ | $\beta$ |
| 1 | Quanta Technology (2009) | 284# | 0.035# |
| 2 | Panteli et al. (2017) | 294# | 0.25# |
| 3 | Fu et al. (2019) | 223.5# | 0.04# |
| 4 | 50th percentile – CO (Odisha, India) | 292.5 | 0.104 |
| 5 | 50th percentile – FD (Odisha, India) | 280.4 | 0.088 |
| # Median and logarithmic standard deviation values of fragility functions are evaluated based on 50th percentile probability, and the difference between the logarithmic values of 50th and 84th percentile probability values, respectively. | | | |

## 6. Conclusions

This paper presents the development of fragility curves for power transmission towers in the state of Odisha, India based on the damages observed during 2019 cyclone Fani. Aleatory and relevant epistemic uncertainties are considered. Key conclusions drawn from the study are as follows.

1. Damage data for transmission towers was obtained from OPTCL. A total of 87 towers had collapsed, while another 41 had received partial damages during cyclone Fani. Two damage states were considered for developing fragility curves: collapse, and functionality disruption. Per the available damage data, 87 towers were in the former damage state, while 128 towers were in the latter.

2. Aleatory uncertainty includes the variabilities in the properties of structural system (e.g., height of tower, material properties), and can be described using the parameters of a lognormal distribution. Epistemic uncertainties associated with estimation of wind speed at the location of towers and the size of damaged sample were considered. While the damage data received from OPTCL was consistent with the track of cyclone Fani, there were reasons to believe that some more towers could have been damaged than that considered in the study. However, the uncertainty associated with the sample size was found to be much smaller compared to that in the estimation of wind speed at a location.

3. Expectedly, the $50^{th}$ percentile fragility curves for collapse were to the right relative to that for functionality disruption (it considers partial damage and collapse). Exceptions are observed at very low probabilities, which is attributed to the fact that both median and logarithmic standard deviation for the $50^{th}$ percentile fragility curve corresponding to collapse are slightly greater than those for the functionality disruption damage state. The $50^{th}$ percentile fragility curves corresponding to the two damage states compared reasonably well with those reported for transmission towers from different parts of the world.

A major limitation of the present study is that the fragility curves were developed using the damage data from a single cyclone. Moreover, the properties of structural system were not available. In spite of these limitations, it is believed that these fragility curves can serve as a benchmark to establish the properties of structural system, loading, and associated uncertainties, which will provide further insights into the influence of structural strengthening on the fragility curves. Such information will be very useful in establishing the optimal strategies to improve the resilience of the power transmission network in the state of Odisha, India. These frameworks can serve as a template for other coastal regions in the Indian subcontinent.

**Acknowledgments**

This research project was financially supported by Indian Institute of Technology Gandhinagar. Authors thank Odisha Power Transmission Corporation Limited (OPTCL) for providing the damage data.

**Appendix A**

*A.1. Radial wind profile models*

Radial wind profile models (RWPM) for tropical cyclones are used to estimate the wind speed at desired locations during cyclonic events. These models yield a wind speed of zero at the eye of the cyclone, which increases sharply to reach the maximum wind speed ($V_{max}$) at the eye wall. The distance between the eye and the location of maximum wind speed is known as the radius of maximum wind speed ($R_{max}$). Beyond eye wall, the wind speed starts decreasing asymptotically to zero away from the eye of the cyclone. In this study, three commonly used RWPMs are considered: Willoughby single exponential (WSE) [18], Willoughby double exponential (WDE) [18], and Holland (HOL) [19]. A brief description of the models is presented in the sections below.

*A.1.1. Willoughby single and double exponential models*

Willoughby single and double exponential models are described as follows.

$$V(r) = V_i = V_{max}\left(\frac{r}{R_{max}}\right)^n, \quad if\ (0 \leq r < R_1) \qquad (A-1)$$

$$V(r) = V_i(1-w) + V_o(w), \text{ if } (R_1 \leq r \leq R_2) \tag{A-2}$$

$$V(r) = V_o = V_{max} \times e^{\left(\frac{-r-R_{max}}{X_1}\right)}, \text{ if } (R_2 < r) \tag{A-3}$$

where, $V(r)$ is gradient-level wind speed at a radial distance $r$ from the eye, $n$ is the power law coefficient governing the increase of wind speed in the zone between eye and the eye wall, $R_1$ and $R_2$ are radial distances from the eye representing the boundaries of the transition zone, $w$ is the weighing factor (takes values between 0 and 1), $X_1$ represents the decay length governing the decay of wind speed after eye wall, and other parameters were defined previously. Parameters $n$, $R_1$, $R_2$, $w$ and $X_1$ are to be evaluated using the equations discussed in [18]. For Willoughby double exponential model, the equation (A-3) is replaced as follows

$$V(r) = V_o = V_{max}\left[(1-A) \times e^{\left(\frac{-r-R_{max}}{X_1}\right)} + A \times e^{\left(\frac{-r-R_{max}}{X_2}\right)}\right], \text{ if } (R_2 < r) \tag{A-4}$$

where, parameter $A$ acts as a proportionating factor between the two exponential decay functions, and other parameters were defined previously.

*A.1.2. Holland model*

Holland model is described as follows.

$$V(r) = \sqrt{\frac{A \times B \times \Delta p \times e^{\left(\frac{-A}{r^B}\right)}}{\rho \times r^B}} \tag{A-5}$$

where, $\Delta p$ is pressure difference between the ambient pressure and the pressure at the eye, $\rho$ is the air density, $A$ and $B$ represent scaling parameters (see [19]), and other parameters were defined previously.

## Appendix B

*B.1. Choice of distribution to characterize fragility curves*

Five bi-parametric distributions were considered to describe the fragility curves: normal (NO), Cauchy (CA), Gamma (GA), Weibull (WE), and lognormal (LO). One hundred damage datasets were generated using the approach presented in Fig. 12, which considers aleatory, and epistemic uncertainties associated with intensity measure and sample size. Least square fit approach was used to establish the parameters of the five bi-parametric distributions for each damage dataset. Fig. B1shows the five distributions fit to a damage dataset. Further, sum of square of errors (SSE) was calculated as follows:

$$\text{SSE} = \sum_{i=1}^{30}\left(y_i - \hat{y}_i\right)^2 \quad \text{(B-1)}$$

where, $i$ is the index of the bin, $y_i$ is the ratio of failed towers to the total number of towers in the bin and $\hat{y}_i$ is the predicted failure ratio of the bin corresponding to its average 3-sec gust speed using the cumulative distribution function fit. Lognormal distribution was associated with least SSE among the five distributions for 98 out of 100 datasets (Fig. B2(a)). The magnitudes of the errors are reported in Fig. B2(b). Accordingly, lognormal distribution is considered most suitable to describe the fragility curves.

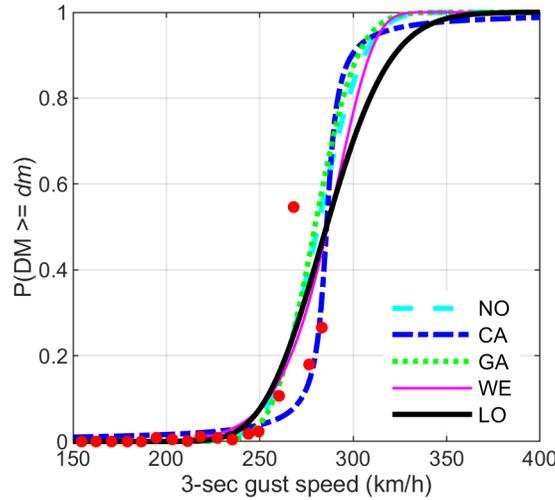

**Fig. B1.** Bi-parametric distributions (Normal – NO, Cauchy – CA, Gamma – GA, Weibull – WE, Lognormal – LO) fit to a dataset.

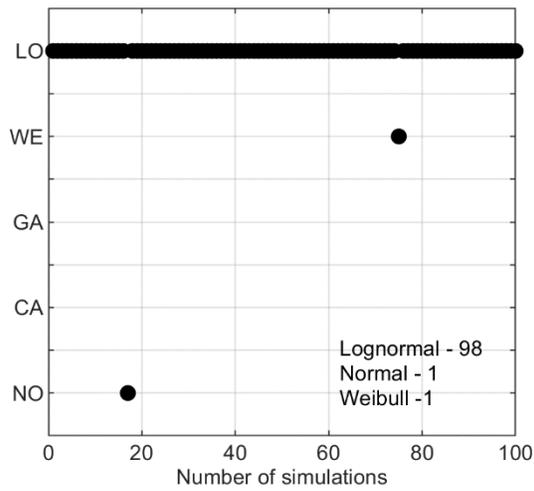 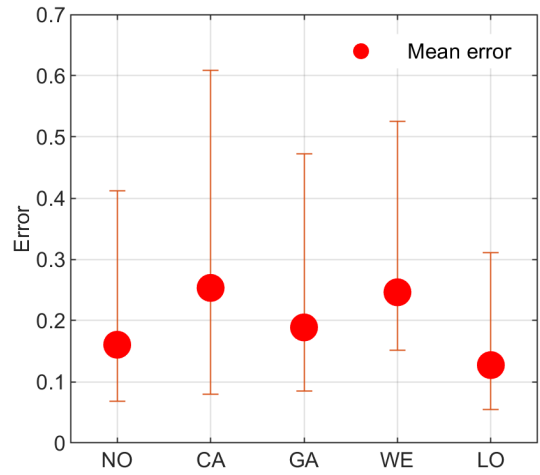

(a) Distributions with minimum SSE

(b) Mean and 68% confidence bounds for estimates of SSEs over 100 simulations

**Fig. B2.** Sum of square of errors corresponding to five bi-parametric distributions.